\documentclass[twocolumn,english,aps,showpacs]{revtex4}
\usepackage[T1]{fontenc}
\usepackage[latin9]{inputenc}
\usepackage[a4paper]{geometry}
\usepackage{babel}
\usepackage{amsmath}
\usepackage{amssymb}
\usepackage{graphicx}
\usepackage{esint}
\usepackage[unicode=true,pdfusetitle,
 bookmarks=true,bookmarksnumbered=false,bookmarksopen=false,
 breaklinks=false,pdfborder={0 0 1},backref=false,colorlinks=false]
 {hyperref}
\usepackage{breakurl}

\makeatletter
\@ifundefined{textcolor}{}
{%
 \definecolor{BLACK}{gray}{0}
 \definecolor{WHITE}{gray}{1}
 \definecolor{RED}{rgb}{1,0,0}
 \definecolor{GREEN}{rgb}{0,1,0}
 \definecolor{BLUE}{rgb}{0,0,1}
 \definecolor{CYAN}{cmyk}{1,0,0,0}
 \definecolor{MAGENTA}{cmyk}{0,1,0,0}
 \definecolor{YELLOW}{cmyk}{0,0,1,0}
}

\usepackage{babel}
\usepackage{babel}
\usepackage{babel}
\usepackage{babel}
\usepackage{babel}
\usepackage{babel}
\usepackage{babel}

\makeatother

\begin{document}

\title{Scaling properties of one-dimensional driven-dissipative condensates}

\author{Liang He$^{1}$, Lukas M. Sieberer$^{1,2}$, Ehud Altman$^{2}$,
and Sebastian Diehl$^{1,3}$}

\affiliation{$^{1}$Institute for Theoretical Physics, University of Innsbruck,
A-6020 Innsbruck, Austria}

\affiliation{$^{2}$Department of Condensed Matter Physics, Weizmann Institute
of Science, Rehovot 7610001, Israel}

\affiliation{$^{3}$Institute for Theoretical Physics, Technical University Dresden,
D-01062 Dresden, Germany}
\begin{abstract}
We numerically investigate the scaling properties of a one-dimensional
driven-dissipative condensate described by a stochastic complex Ginzburg-Landau
equation (SCGLE). We directly extract the static and dynamical scaling
exponents from the dynamics of the condensate's phase field, and find
that both coincide with the ones of the one-dimensional Kardar-Parisi-Zhang
(KPZ) equation. We furthermore calculate the spatial and the temporal
two-point correlation functions of the condensate field itself. The
decay of the temporal two-point correlator assumes a stretched-exponential
form, providing further quantitative evidence for an effective KPZ
description. Moreover, we confirm the observability of this non-equilibrium
scaling for typical current experimental setups with exciton-polariton
systems, if cavities with a reduced $Q$ factor are used. 
\end{abstract}


\maketitle

\section{Introduction}

Physical systems driven far away from thermal equilibrium can show
intrinsically different properties from their equilibrium counterparts.
One prototypical example is the growing interface, whose long-wavelength
dynamics, described by the so-called Kardar-Parisi-Zhang (KPZ) equation
\cite{KPZ_equation}, does not belong to the Halperin-Hohenberg classification
of near thermal equilibrium dynamical behavior \cite{HH_models}.
Recent experimental progress in realizing Bose-Einstein condensation
of exciton-polaritons in pumped semiconductor heterostructures \cite{Polariton_Experiment_1,Polariton_Experiment_2,Polariton_Experiment_3}
holds the promise of developing such systems into laboratories for
non-equilibrium statistical mechanics. Microscopically, these systems
exhibit coherent and driven-dissipative dynamics on an equal footing,
and therefore explicitly violate detailed balance characteristic of
an equilibrium system. This phenomenology can have drastic consequences
for the macrophysics of such systems. Indeed, for the case of driven-dissipative
condensates of exciton-polaritons in two dimensions (2D), it was pointed
out recently \cite{Altman_2d_driven_SF_2013} that quasi-long-range
order can not exist in the long-wavelength limit, in stark contrast
to the familiar properties of 2D equilibrium condensates. This conclusion
was drawn from a connection between the stochastic complex Ginzburg-Landau
equation (SCGLE) and the KPZ equation in the long-wave length limit
\cite{Grinstein_KPZ_SCGLE,Altman_2d_driven_SF_2013}, as also noticed
in 1D \cite{Wouters_1D_static}. However, direct numerical evidence
of this connection is still missing.

As a first step to fill this gap, here we investigate the long-wavelength
behavior of the dynamics of a driven-dissipative condensate in 1D.
Our first goal is to study whether scaling properties of the condensate's
phase field dynamics, in particular static and dynamical exponents,
indeed coincide with those implied by the KPZ equation. The second
goal is to directly study both the spatial and temporal correlation
function of the bosonic field for the condensate itself to see whether
they match the prediction from the KPZ picture. We note that a similar
investigation is reported in \cite{Wouters_1d_temproal}, and comment
on the relation to the present work in Sec. IV.

We achieve our goals via direct numerical simulations of the SCGLE
which governs the dynamics of driven-dissipative condensates. We directly
extract both the static and dynamical critical exponents of the system
from the dynamics of the condensate's phase field. Within numerical
error, we indeed find that the critical exponents of the SCGLE coincide
with the ones of the KPZ equation (see Figs.~\ref{Fig. w_scaling_collapse},
\ref{Fig. roughness_exponent}, and \ref{Fig. growth_exponnet}),
and we estimate the crossover time scale (see Fig. \ref{Fig. Crossover_time_scale})
beyond which the KPZ scaling behavior can be observed. We further
find that the scaling properties of the condensate field dynamics
(see Figs.~\ref{Fig. Psi_Psi_spatial_correlation} and \ref{Fig. Psi_Psi_temporal_correlation_LogLog_Log_scale_plot})
match the expectation from the effective description in terms of the
KPZ equation. Finally, we demonstrate that the KPZ scaling can be
seen in current experimental setups with exciton-polaritons, if cavities
with a reduced $Q$ factor are used (see Fig.~\ref{Fig. Psi_Psi_temporal_correlation_LogLog_Log_scale_plot}).

Our numerical approach is based on an effective lower polariton dynamical
model with a quartic nonlinearity. Another widely used dynamical model
for exciton-polariton condensates is the so-called generalized Gross-Pitaevskii
equation \cite{Carusotto2013,Wouters_1D_static,Wouters_1d_temproal},
which results from a two-band model after tracing out the reservoir
band \cite{Carusotto2013}. As a matter of fact these two models coincide
with each other after expanding the nonlinear term in the generalized
Gross-Pitaevskii equation with respect to the polariton field (see
for instance Eq. (\ref{eq:Exp_rescaled_SCGLE})). We note that KPZ
scaling behavior only emerges at long wavelengths, where general power
counting arguments ensure that this approximation (with low order
in the polariton fields) is well justified in the long wavelength
limit.

The paper is organized as follows: In Sec. II, we specify the system
and model under study, and the theoretical approach used. In Sec.
III, we present a detailed discussion of the scaling properties of
the phase field correlations. This contains in particular the static
and dynamical exponents of the condensate's phase field dynamics.
In Sec. IV, we discuss the scaling properties of two-point correlation
functions of the condensate field itself. In Sec. V, we investigate
the experimental observability of the scaling properties discussed
in Sec. IV in exciton-polariton condensate experiments. We conclude
and give an outlook in Sec. VI.

\section{Model and Theoretical Approach}

The dynamics of driven-dissipative condensates, which have been realized
in experiments with exciton-polariton systems \cite{Polariton_Experiment_2,Polariton_Experiment_3,Polariton_Experiment_1},
can be modeled by the SCGLE with a complex Gaussian white noise (units
are chosen such that $\hbar=1$) which reads in 1D as \cite{Carusotto2013,Altman_2d_driven_SF_2013}
\begin{equation}
\frac{\partial}{\partial\tilde{t}}\tilde{\psi}=\left[\tilde{r}+\tilde{K}\frac{\partial^{2}}{\partial\tilde{x}^{2}}+\tilde{u}|\tilde{\psi}|^{2}\right]\tilde{\psi}+\tilde{\zeta}\label{eq:SCGLE}
\end{equation}
with $\tilde{r}=-\tilde{r}_{d}-i\tilde{r}_{c}$, $\tilde{K}=\tilde{K}_{d}+i\tilde{K}_{c}$,
$\tilde{u}=-\tilde{u}_{d}-i\tilde{u}_{c}$, $\langle\tilde{\zeta}(\tilde{x},\tilde{t})\tilde{\zeta}(\tilde{x}',\tilde{t}')\rangle=0$,
$\langle\tilde{\zeta}^{*}(\tilde{x},\tilde{t})\tilde{\zeta}(\tilde{x}',\tilde{t}')\rangle=2\tilde{\sigma}\delta(\tilde{x}-\tilde{x}')\delta(\tilde{t}-\tilde{t}')$.
The second moment of the noise $\tilde{\sigma}=\tilde{\gamma}_{l}$
with $\tilde{\gamma}_{l}$ being the single particle loss, while $\tilde{r}_{d}=\tilde{\gamma}_{l}-\tilde{\gamma}_{p}$
is the difference between the single particle loss and pump. For the
existence of a condensate in the mean field steady state solution,
$\tilde{r}_{d}$ has to be negative, i.e., the single-particle pump
rate has to be larger than the loss rate. $\tilde{u}_{d}$ is the
positive two-particle loss rate; $K_{c}=1/(2m_{\mathrm{LP}})$ with
$m_{\mathrm{LP}}$ being the mass of polaritons and $K_{d}$ is an
effective diffusion constant. For convenience, we use the following
rescaled form of Eq. (\ref{eq:SCGLE}), 
\begin{eqnarray}
\frac{\partial}{\partial t}\psi & = & \left[r+K\frac{\partial^{2}}{\partial x^{2}}+u|\psi|^{2}\right]\psi+\zeta,\label{eq:Dimension_less_SCGLE}
\end{eqnarray}
where

\begin{eqnarray}
 &  & t=|\tilde{r}_{d}|\tilde{t},\, x=\sqrt{\frac{|\tilde{r}_{d}|}{\tilde{K}_{d}}}\tilde{x},\\
 &  & \psi=\sqrt{\frac{\tilde{u}_{d}}{|\tilde{r}_{d}|}}\tilde{\psi},\,\zeta=\sqrt{\frac{\tilde{u}_{d}}{|\tilde{r}_{d}|^{3}}}\tilde{\zeta},\\
 &  & r_{c}=\frac{\tilde{r}_{c}}{|\tilde{r}_{d}|},\, K_{c}=\frac{\tilde{K}_{c}}{\tilde{K}_{d}},\, u_{c}=\frac{\tilde{u}_{c}}{\tilde{u}_{d}},\\
 &  & r=1-ir_{c},\, K=1+iK_{c},\, u=-1-iu_{c},
\end{eqnarray}
and the second moment of the rescaled Gaussian white noise $\zeta(x,t)$
is $\sigma=\tilde{\sigma}\tilde{u}_{d}\left|\tilde{r}_{d}\right|^{-3/2}\tilde{K}_{d}^{-1/2}$.

Adopting the amplitude-phase representation of the complex bosonic
field $\psi(x,t)=\rho(x,t)e^{i\theta(x,t)}$, it was shown \cite{Grinstein_KPZ_SCGLE,Altman_2d_driven_SF_2013,Wouters_1D_static}
that, assuming that spatial-temporal fluctuations of the amplitude
field $\rho(x,t)$ are small, the dynamical equation of the phase
field $\theta(x,t)$ assumes in the low-frequency and long-wavelength
limit the form of the KPZ equation, which reads 
\begin{equation}
\partial_{t}\theta(x,t)=D\partial_{x}^{2}\theta(x,t)+\frac{\lambda}{2}\left(\partial_{x}\theta(x,t)\right)^{2}+\eta(x,t),\label{eq:KPZ_equation}
\end{equation}
where $\eta(x,t)$ is an effective Gaussian white noise, with mean
$\langle\eta(x,t)\rangle=0$, and correlations $\langle\eta(x,t)\eta(x',t')\rangle=2\sigma_{\mathrm{KPZ}}\delta(x-x')\delta(t-t')$.
Here $\sigma_{\mathrm{KPZ}}=(\tilde{u}_{d}^{2}+\tilde{u}_{c}^{2})\tilde{\gamma}_{l}/(2\tilde{u}_{d}(\tilde{\gamma}_{p}-\tilde{\gamma}_{l}))$
is the effective noise strength, $D=\tilde{K}_{d}(1+\tilde{K}_{c}\tilde{u}_{c}/\tilde{K}_{d}\tilde{u}_{d})$
is the diffusion constant, and $\lambda=2\tilde{K}_{c}\left(\tilde{K}_{d}\tilde{u}_{c}/\tilde{K}_{c}\tilde{u}_{d}-1\right)$
is the non-linear coupling strength \cite{Altman_2d_driven_SF_2013}.
With a simple rescaling, i.e., $\theta=\Theta\sqrt{2\sigma_{\mathrm{KPZ}}/D},\, t=\tau/D,\,\eta=\xi\sqrt{2\sigma_{\mathrm{KPZ}}D}$,
the KPZ equation Eq. (\ref{eq:KPZ_equation}) can be recast into a
form where only one dimensionless parameter, the non-linear coupling
strength $g$, enters, i.e. 
\begin{equation}
\partial_{\tau}\Theta(x,\tau)=\partial_{x}^{2}\Theta(x,\tau)+g\left(\partial_{x}\Theta(x,\tau)\right)^{2}+\xi(x,\tau),\label{eq:rescaled_KPZ}
\end{equation}
where 
\begin{equation}
g=\lambda\sqrt{\frac{\sigma_{\mathrm{KPZ}}}{2D^{3}}},\label{eq:g_non-equilibrium strength}
\end{equation}
and $\langle\xi(x,\tau)\xi(x',\tau')\rangle=\delta(x-x')\delta(\tau-\tau')$.
Importantly, the magnitude of $g$ directly characterizes how far
the dynamics of the complex field $\psi$ is driven away from thermal
equilibrium. More precisely, $g=0$ is guaranteed by symmetry in a
thermal equilibrium system which obeys global detailed balance \cite{Sieberer_PRL_PRB},
in which case Eq. (\ref{eq:rescaled_KPZ}) reduces to the so-called
Edwards-Wilkinson (EW) dynamical equation \cite{EW-Dynamics}, while
$g\neq0$ indicates that the system is driven away from thermal equilibrium.

In the following, we investigate the scaling properties of various
correlation functions of the phase field $\theta(x,t)$, in particular
the static and dynamical critical exponent, as well as the correlation
properties of the complex bosonic field $\psi(x,t)$ which are of
most direct physical interest for experiments.

To put our investigation in a more general context, here we mention
a few situations where similar dynamical equations appear. Without
the noise term in (\ref{eq:Dimension_less_SCGLE}), the above equation
reduces to the \emph{deterministic} complex Ginzburg-Landau equation
(CGLE). One key feature of the latter is the existence of a so-called
Benjamin-Feir unstable parameter region \cite{Benjamin_Feir_region}
specified by $1+K_{c}u_{c}<0$, where the dynamics described by the
deterministic CGLE develops spatiotemporal chaotic behavior (see e.g.
\cite{Grinstein_phase_turbulence_in_CGLE}) which has been extensively
studied in the literature \cite{Cross_Hohenberg_Pattern_formation,Aranson_CGLE}.
As we are interested in the parameter regime with both $K_{c}$ and
$u_{c}$ being positive, the Benjamin-Feir unstable region is not
relevant for the current investigation. However, it can be relevant
if one is interested in turbulence of the bosonic fluid in the presence
of external noise \cite{Gasenzer_noisy_turbulence}. Moreover, a similar
stochastic dynamical equation, the so-called stochastic Gross-Pitaevskii
equation \cite{Stoof_1999,Blakie_Gardiner_SCGLE}, is used to describe,
e.g. the BEC formation dynamics of alkali atoms at finite temperature.
Here, however, the constraints resulting from detailed balance in
stationary state are built in. Finally, we mention that recently in
Ref. \cite{Wouters_1D_static} a higher order spatial derivative term
was included in the effective description of the 1D SCGLE. This study
focuses on the static correlation properties of the system, where
a crossover in the spatial correlation function at intermediate scale
is identified.

We finally give some general information concerning our numerical
simulations. We use the semi-implicit algorithm developed in \cite{Drummond_1997}
to solve the stochastic partial differential equation (\ref{eq:Dimension_less_SCGLE})
numerically. In all the simulations spatial periodic boundary conditions
of the complex field $\psi(x,t)$ are assumed and the winding number
of the phase field $\theta(x,t)$ across the whole system is chosen
to be zero. We work in the low noise regime, where we find defects
of the phase field to be absent. If not specified in text, we use
$N_{\mathrm{Traj}}=10^{2}$ stochastic trajectories to perform ensemble
averages.

\section{Scaling properties of the phase correlations \label{sec:scaling_properties_from_phase_field}}

\subsection{KPZ exponents}

In order to characterize the phase dynamics we extract the phase field
$\theta(x,t)$ from the simulations of the condensate field $\psi(x,t)$.
We then investigate the following correlation function associated
with the phase field: 
\begin{eqnarray}
w(L,t) & \equiv & \left\langle \frac{1}{L}\int_{x}\theta^{2}(x,t)-\left(\frac{1}{L}\int_{x}\theta(x,t)\right)^{2}\right\rangle ,\label{eq:Roughness_function_definition}
\end{eqnarray}
where $L$ is the linear size of the system and ``$\langle\,\rangle$''
indicates ensemble average over stochastic trajectories. In the context
of the KPZ equation, $w(L,t)$ is usually referred to as ``roughness
function''. Regarding $\theta(x,t)$ as the crystal height variable
as in the conventional KPZ equation, $w(L,t)$ measures the spatial
fluctuations of that height. Later we discuss subtleties involved
in the definition of $w(L,t)$ due to the fact that the phase field
$\theta$ is in fact compact, i.e. defined on the circle. Measuring
the scaling properties of $w(L,t)$ allows to extract both static
and dynamic exponents and thus establish a connection to KPZ universality
(see e.g. \cite{Marinari_KPZ_Numerics}): 
\begin{enumerate}
\item \label{item:growth_exponent}In a large system we expect to see a
wide range of time-scales over which $w(L,t)\propto t^{2\beta}$,
where the dynamical exponent $\beta$ is usually referred to as growth
exponent in the KPZ context. It relates to the conventional dynamical
exponent $z$ according to $\beta=\alpha/z$, with $\alpha$ being
the roughness exponent to be explained in the following. 
\item \label{item:static_exponent}Because of the finite system size, the
roughness function will saturate at $w_{s}(L)$ beyond a saturation
time. We expect the saturation value to scale as $w_{s}(L)\sim L^{2\alpha}$
, where the static exponent $\alpha$ is called the roughness exponent
in the KPZ context. 
\item The roughness function reaches its saturation value $w_{s}(L)$ at
a time $T_{s}$, which thus separates the growth period \ref{item:growth_exponent}.
from the long time regime \ref{item:static_exponent}. This saturation
time scales with system size as $T_{s}\sim L^{z}$. 
\end{enumerate}
\begin{figure}
\includegraphics[width=3.5in]{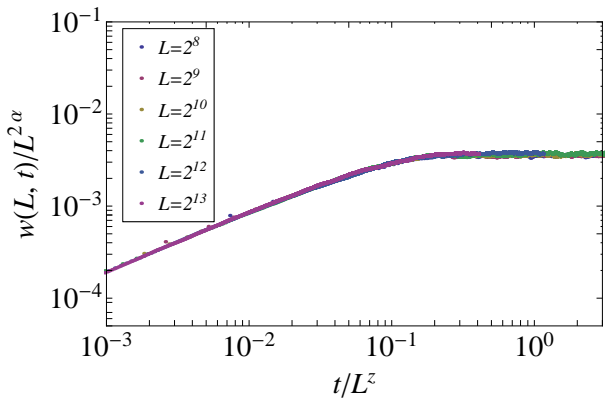}

\protect\protect\caption{(Color online) Finite-size scaling collapse of $w(L,t)$ in the 1D
KPZ universality class with $\alpha=1/2$ and $z=3/2$. Ensemble averages
were performed over a number of $N_{\mathrm{Traj}}=1000$ stochastic
trajectories. Values of the parameters used in the simulations are
$r_{c}=-0.1,u_{c}=0.1,\sigma=0.1,K_{c}=3.0$.}

\label{Fig. w_scaling_collapse} 
\end{figure}

These scaling features are demonstrated by the finite-size scaling
of $w(L,t)$ shown in Fig.~\ref{Fig. w_scaling_collapse}. Perfect
data collapse is obtained using the 1D KPZ exponents $\alpha=1/2$
and $z=3/2$. During the growth period the roughness increases nearly
linearly on the log-log scale, which indicates power-law growth $w(L,t)\propto t^{2\beta}$.
For different system sizes saturation is reached at the same point
on the rescaled time axis, confirming the scaling behavior $T_{s}\sim L^{z}$.
Finally, the saturation values $w_{s}(L)$ of the roughness function
collapse upon rescaling $w(L,t)$ with $L^{2\alpha}$.

A more precise numerical determination of the exponents $\alpha$
and $\beta$, which confirms that their values are given by the ones
of the KPZ equation, i.e. $\alpha=1/2$ and $\beta=1/3$, is presented
in the appendix. This provides us with strong evidence that the phase
field dynamics of a driven-dissipative condensate is indeed described
by the KPZ equation, in contrast to the thermal equilibrium case,
in which the dynamics of the phase is purely diffusive and thus belongs
to the EW universality class \cite{Halpin-Healy_KPZ_review}. The
corresponding dynamical exponent $\beta=1/4$ is different from KPZ
universality, however, the value of the static roughness exponent,
$\alpha=1/2$, is exactly the same in both cases. This is due to a
symmetry of the KPZ equation that is present only in one spatial dimension,
and which allows one to show that the static correlations in stationary
state are Gaussian \cite{Taeuber_book}. On the other hand, the dynamical
exponent $\beta$ (or equivalently $z$) witnesses quantitatively
the difference between KPZ and EW universality.

Before we proceed, let us emphasize an important difference between
the phase of a complex field we consider here and the crystal height:
the phase is a compact field variable defined on a circle. Without
loss of generality the value of $\theta(x,t)$ is in fact bounded
to the interval $(-\pi,\pi]$. Consequently, the value of $w(L,t)$
is also bounded from above by $4\pi^{2}$, which inevitably invalidates
the static scaling behavior $w_{s}(L)\sim L^{2\alpha}$ if $\alpha$
is positive as expected from the conventional KPZ scenario. However,
as long as the field amplitude remains nonvanishing we can let the
value of $\psi$ be defined on the Riemann surface, where the value
of $\theta$ is in the interval $(-\infty,+\infty)$. With this choice
there is no upper bound imposed on $w_{s}(L)$. In numerical simulations,
we ensure the requirement $|\psi(x,t)|>0$ by working with low noise.
In this regime phase defects do not occur within the spatio-temporal
range of our simulations. $\theta(x,t)$ is constructed from $\psi(x,t)$'s
complex argument by requiring the phase difference between neighboring
space-time points to be less than $\pi$.

\subsection{Crossover time scale\label{sub:Crossover-time-scale}}

\begin{figure}
\includegraphics[width=2.9in]{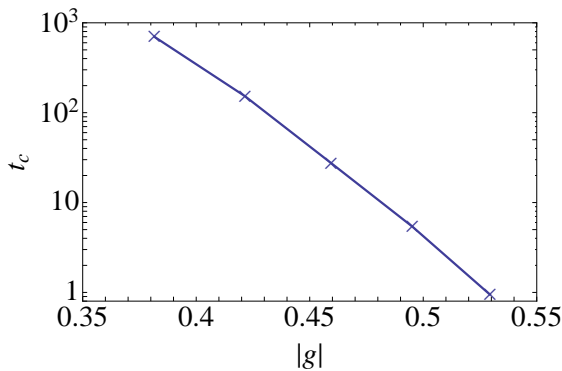}\protect\protect\caption{(Color online) Dependence of the crossover time $t_{c}$ on the non-equilibrium
strength $|g|$. $t_{c}$ decreases pronouncedly as $|g|$ increases.
In the numerical results presented here, $|g|$ is tuned by changing
$K_{c}=2.4,2.2,2.0.1.8,1.6$ while keeping the other parameters unchanged.
Their values are $L=2^{15},r_{c}=-0.1,u_{c}=0.1,\sigma=0.1$.}

\label{Fig. Crossover_time_scale} 
\end{figure}

In the above subsection we have established that the phase field dynamics
indeed belong to the KPZ universality class. However, it is important
to notice that the scaling behavior of $w(L,t)\propto t^{2\beta}$,
where $\beta=1/3$ is the KPZ growth exponent, is reached only after
a crossover time $t_{c}$. In particular, for weak nonlinearity (i.e.
$|g|\ll1$) the KPZ renormalization group equations lead to a crossover
time that scales as $t_{c}\approx t_{0}|g|^{-4}$ \cite{Natermann_1d_cross_over_time_scaling},
where $t_{0}$ is a microscopic time scale. Scaling behavior of $w(L,t)$
before $t_{c}$ is expected to be governed by the EW growth exponent
$\beta=1/4$. In Fig.~\ref{Fig. Crossover_time_scale}, we investigate
the $|g|$ dependence of the crossover time $t_{c}$ at moderate values
of $|g|$ (the numerical scheme for the extraction of $t_{c}$ can
be found in App.~\ref{sec:dynam-growth-expon}), since extraction
of $t_{c}$ in the near equilibrium case, $|g|\ll1$, is numerically
very demanding and $t_{c}$ quickly exceeds the accessible simulation
runtimes. We observe that $t_{c}$ increases pronouncedly as $|g|$
decreases (but not yet according to the weak coupling scaling pointed
out above). The rapid decrease of $t_{c}$ with increasing non-equilibrium
strength is promising for the experimental observation of KPZ scaling
behavior, rather than transient EW-like dynamics, before finite size
effects set in. We discuss possible experimental settings for observing
these phenomena in Sec.~\ref{sec:Experiments_related_discussion}.

\section{Scaling of the condensate field correlations}

In the previous section we have demonstrated numerically that the
dynamics of the phase of a one-dimensional polariton condensate follows
universal KPZ scaling. In this section we investigate how this scaling
manifests in directly observable correlations of the condensate field.
Specifically, we consider the correlation functions

\begin{eqnarray}
C_{x}(x_{1},x_{2};t) & \equiv & \langle\psi^{*}(x_{1},t)\psi(x_{2},t)\rangle,\\
C_{t}(x;t_{1},t_{2}) & \equiv & \langle\psi^{*}(x,t_{1})\psi(x,t_{2})\rangle,
\end{eqnarray}
i.e. the equal time two-point correlation function in space and the
temporal autocorrelation function, respectively. These are directly
accessible in experiments with exciton-polaritons: Both spatial and
temporal coherence can be probed by interference measurements, on
the photoluminescence emitted from different regions of the exciton-polariton
condensate \cite{Polariton_Experiment_1,Polariton_Experiment_3} and
by combining two images of the condensate taken at different times
using, e.g., a Mach-Zehnder interferometer \cite{exp_g1_fun_measure},
respectively. The visibility of interference fringes yields the correlation
functions. Assuming spatial translational invariance of the correlation
functions, we calculate the following spatially averaged correlation
functions, which are equivalent to the corresponding correlation functions
above but in practice help to reduce the statistical error, 
\begin{equation}
\begin{split}\bar{C}_{x}(x_{1},x_{2},t) & \equiv\frac{1}{L}\int dy\langle\psi^{*}(x_{1}+y,t)\psi(x_{2}+y,t)\rangle,\\
\bar{C}_{t}(t_{1},t_{2}) & \equiv\frac{1}{L}\int dx\langle\psi^{*}(x,t_{1})\psi(x,t_{2})\rangle.
\end{split}
\end{equation}

\subsection{Spatial correlations }

\begin{figure}
\includegraphics[width=3in]{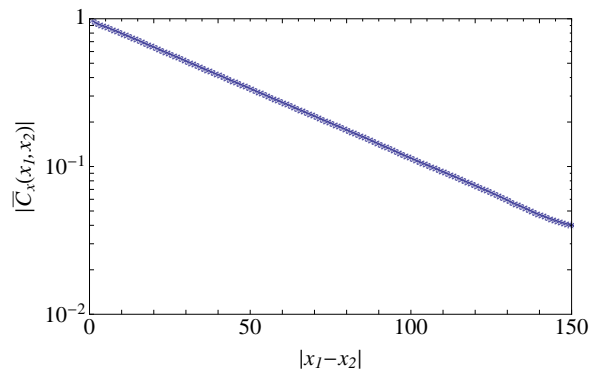}\protect\protect\caption{(Color online) Behavior of the translation invariant two-point function
$\bar{C}_{x}(x_{1},x_{2},t=2.9\times10^{5}>T_{s})$ at linear system
size $L=2^{12}$ on a semi-logarithmic scale. $N_{\mathrm{Traj}}=800$
stochastic trajectories are used to perform the ensemble average.
Values of other parameters used here are $r_{c}=-0.1,u_{c}=0.1,\sigma=0.1,K_{c}=3.0$.
The exponential decay of $\left|\bar{C}_{x}(x_{1},x_{2},t)\right|$
with respect to $|x_{1}-x_{2}|$ can be clearly identified from this
plot. }

\label{Fig. Psi_Psi_spatial_correlation} 
\end{figure}

We start with the spatial correlation function $\bar{C}_{x}(x_{1},x_{2},t)$.
In Fig. \ref{Fig. Psi_Psi_spatial_correlation} we show the dependence
of $\left|\bar{C}_{x}(x_{1},x_{2},t)\right|$ on the distance $|x_{1}-x_{2}|$
at time $t>T_{s}$, from which we clearly identify exponential decay
on the semi-logarithmic scale plot. This coincides with the prediction
from the effective KPZ description in 1D, and with previous numerical
results \cite{Wouters_1D_static}. However, as anticipated in Sec.
\ref{sec:scaling_properties_from_phase_field}, this static signature
would in fact be compatible with thermal equilibrium dynamics of the
field $\psi$ and does not unambiguously demonstrate KPZ physics.

\subsection{Temporal correlations\label{sub:Temporal-correlations}}

\begin{figure}
\includegraphics[width=3.3in]{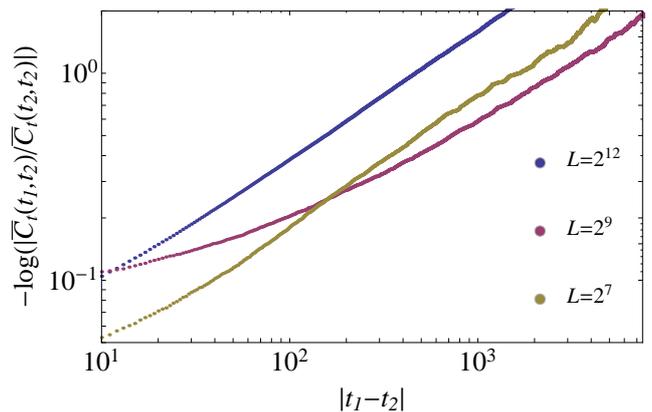}

\protect\protect\caption{(Color online) The dependence of $-\log\left(\left|\bar{C}_{t}(t_{1},t_{2})\right|/\left|\bar{C}_{t}(t_{2},t_{2})\right|\right)$
on $|t_{1}-t_{2}|$ for three different sets of parameters in the
SCGLE and system sizes. The system size and the parameters for the
uppermost (blue) curve are the same as those used in Fig.~\ref{Fig. Psi_Psi_spatial_correlation}.
For the lowermost (at large time differences) curve shown in red,
the dimensionless linear system size is $2^{9},$ and the parameters
are chosen to match typical values in current experiments with exciton-polaritons
(see Sec.~\ref{sec:Experiments_related_discussion} for details).
Finally, assuming that a cavity with reduced $Q$ factor is used we
obtain the parameters for the middle (yellow) curve, which corresponds
to a system size of $2^{7}$. KPZ behavior is revealed by performing
linear fits to the data points: with $|t_{1}-t_{2}|\in[10^{2},10^{3}]$
we find $\beta=0.311$ and $\beta=0.317$ for the blue and yellow
curves, respectively, while for the red curve a fit with $|t_{1}-t_{2}|$
lying in the last half decade in the above plot, gives rise to $\beta=0.307$.
These values should be compared with the KPZ prediction $\beta=1/3$.
For all curves $N_{\mathrm{Traj}}=10^{3}$ stochastic trajectories
are used.}

\label{Fig. Psi_Psi_temporal_correlation_LogLog_Log_scale_plot} 
\end{figure}

In contrast to the time-independent spatial correlation function discussed
in the previous section, the temporal correlation function shows distinct
properties depending on whether the system is in thermal equilibrium
or not: indeed, based on the effective long-wavelength description
of the out-of-equilibrium condensate dynamics in terms of the KPZ
equation, we expect stretched-exponential decay of the autocorrelation
function, i.e., $\left|\bar{C}_{t}(t_{1},t_{2})\right|=Ae^{-B|t_{1}-t_{2}|{}^{2\beta}}$,
with the KPZ growth exponent $\beta=1/3$ and non-universal numbers
$A$ and $B$. On the other hand, the purely diffusive EW dynamics
of the phase of a condensate in equilibrium entails decay with an
exponent $\beta=1/4$. Hence both cases lead to linear growth of $-\log\left(\left|\bar{C}_{t}(t_{1},t_{2})\right|/\left|\bar{C}_{t}(t_{2},t_{2})\right|\right)$
with a slope of $2\beta$ in the double-logarithmic scale used in
Fig. \ref{Fig. Psi_Psi_temporal_correlation_LogLog_Log_scale_plot},
which is clearly visible for the upper (at large $|t_{1}-t_{2}|$)
curves shown in blue and yellow. Performing linear fits to the data
points with $|t_{1}-t_{2}|\in[10^{2},10^{3}]$ we find $\beta=0.311$
and $\beta=0.317$, respectively, in reasonable agreement with the
KPZ prediction of $\beta=1/3$ and evidently distinct from the value
$\beta=1/4$ for a condensate in equilibrium. For these curves KPZ
scaling sets in after a short crossover time difference $t_{c}$,
which is due to the relatively large value of the effective non-linear
coupling strength $|g|$ in both cases. On the contrary, for the parameters
that yield the lowermost (red) curve, the value of $|g|$ is small,
and as a result in this case universal scaling behavior is approached
only at the largest time differences shown. A fit with $|t_{1}-t_{2}|$
lying in the last half decade of the data shown in the Figure gives
$\beta=0.307$, and we expect a value closer to $\beta=1/3$ at time
differences larger than those that are accessible within the temporal
range of our simulations. The parameters leading to the two lower
(red and yellow) curves shown in Fig. \ref{Fig. Psi_Psi_temporal_correlation_LogLog_Log_scale_plot}
are relevant for current experiments with exciton-polaritons as is
discussed in the following section.

We note that in a recent work by K. Ji et al. \cite{Wouters_1d_temproal},
the scaling properties of temporal correlation functions of exciton-polariton
condensate were also investigated and among other results, the dynamical
exponent $z$ was extracted. For the case of an exciton-polariton
condensate without elastic collisions ($\tilde{u}_{c}=0$) and $\tilde{K}_{d}=0$,
indicating an infinitely large KPZ nonlinearity $|g|$, they identified
KPZ scaling with $z=3/2$ from simulations of a system with dimensionless
size $L=2^{7}$ \cite{Wouters_1d_temproal}. Instead, for the case
of an exciton-polariton condensate with elastic collisions, indicating
a finite KPZ nonlinearity $|g|$, they reported a dynamical exponent
$z\approx1.7$. This is equivalent to a growth exponent $\beta=\alpha/z=0.5/1.7\simeq0.294$.
As pointed out in \cite{Wouters_1d_temproal}, this deviation from
KPZ scaling could be attributed to finite size effects for the particular
parameter choice in this case. 

This is compatible with our findings: in fact, in order to unambiguously
reveal KPZ scaling in generic cases, relevant to experiments of exciton-polariton
condensates, one needs to simulate large enough systems. Here we have
simulated systems with sizes up to $L=2^{17}$ and confirm from both
the phase correlations (cf. Sec. \ref{sec:scaling_properties_from_phase_field}
and App. \ref{sec:dynam-growth-expon}) and the condensate field correlations
that the growth exponent $\beta$ indeed approaches the expected universal
value for generic parameter values. In the following section, based
on these findings we study whether this asymptotic behavior can be
identified for realistic system parameters.

\section{Predictions for experimental observation\label{sec:Experiments_related_discussion}}

In the preceding sections we studied the SCGLE as an effective description
of the long-wavelength dynamics of a generic driven-dissipative condensate.
The microscopic model for the specific case of exciton-polaritons~\cite{Carusotto2013}
differs from the SCGLE in that the diffusion constant is essentially
absent and instead of an explicit two-body loss term the pump itself
is assumed to be non-linear and saturates at high densities. Slightly
above the condensation threshold the saturable pump term can be expanded
in the polariton field and we recover the SCGLE, which then reads
in dimensionless form 
\begin{equation}
\partial_{t}\psi=\left[i\partial_{x}^{2}+(i+u_{d})\left(\frac{p}{u_{d}}-|\psi|^{2}\right)\right]\psi+\zeta.\label{eq:Exp_rescaled_SCGLE}
\end{equation}
Here the effective dimensionless two-body loss coefficient $u_{d}$
and the dimensionless pump strength $p$ are given by 
\begin{equation}
u_{d}=\frac{\hbar\tilde{\gamma}_{l}R}{2\gamma_{R}\tilde{u}_{c}}(1+2p),\, p=\frac{1}{2}\left(\frac{P}{P_{\mathrm{th}}}-1\right),\label{eq:params_Exp_rescaled_SCGLE}
\end{equation}
with $P$ and $P_{\mathrm{th}}=\tilde{\gamma}_{l}\gamma_{R}/R$ being
the pump rate of the excitonic reservoir and its value at threshold,
respectively; $R$ is the condensate amplification rate and $\gamma_{R}$
denotes the relaxation rate of the reservoir. Finally, $\tilde{\gamma}_{l}$
is the inverse lifetime of polaritons and $\tilde{u}_{c}$ their interaction
strength. Here we measure time and space in units of $\tilde{\gamma}_{l}^{-1}$
and $\sqrt{\hbar/2m_{\mathrm{LP}}\tilde{\gamma}_{l}}$ respectively
with $m_{\mathrm{LP}}$ being the effective mass of lower polaritons.
The strength of the dimensionless noise field $\zeta$ is $\sigma=\tilde{u}_{c}\sqrt{2m_{\mathrm{LP}}/\hbar^{3}\tilde{\gamma}_{l}}$.
Typical values of experimental parameters in 1D exciton-polariton
systems are (see, e.g., Ref. \cite{1D_experiment}), 
\begin{equation}
\begin{array}{l}
m_{\mathrm{LP}}=4\times10^{-5}m_{e},\,\tilde{u}_{c}=5\times10^{-4}\mathrm{meV}\mu\mathrm{m},\\
\tilde{\gamma}_{l}=0.03\mathrm{ps}^{-1},\, R=3\mu\mathrm{m}\cdot\mathrm{ps}^{-1},\,\gamma_{R}=0.06\mathrm{ps}^{-1},
\end{array}\label{eq:Exp_parameters_values_decent_cavity}
\end{equation}
where $m_{e}$ is the mass of the electron.

The lowermost (red) curve in Fig.~\ref{Fig. Psi_Psi_temporal_correlation_LogLog_Log_scale_plot}
shows the temporal correlation function $\bar{C}_{t}(t_{1},t_{2})$
in the stationary state for the values given in Eq. (\ref{eq:Exp_parameters_values_decent_cavity})
and at a dimensionless pump power of $p=0.3$. Due to fact that the
corresponding $|g|$ is relatively small, the red curve approaches
linear growth characteristic of KPZ scaling only after a large crossover
time difference $t_{c}$. As already mentioned in the previous section,
a linear fit to the data points with $|t_{1}-t_{2}|$ lying in the
last half decade in Fig.~\ref{Fig. Psi_Psi_temporal_correlation_LogLog_Log_scale_plot}
yields $\beta=0.307$, indicating that signatures of KPZ physics are
nevertheless observable. However, we note that the physical system
size corresponding to the dimensionless linear system size of $L=2^{9}$
chosen in this simulation is $\sim3\times10^{3}\mu\mathrm{m}$, which
is considerably larger than the typical scale $\sim10^{2}\mu\mathrm{m}$
of current experiments.

Here we propose to make the KPZ physics observable with current experimental
system sizes by reducing the cavity $Q$ factor. To this end, we note
that KPZ scaling is still observable when, while reducing the \emph{physical}
system size, the \emph{dimensionless effective system} size can be
kept large. A convenient knob to achieve this goal is indeed a reduction
of the cavity $Q$ (and thus increase of the decay rate $\tilde{\gamma}_{l}$),
which leads to a decrease of the unit of length. (We note that this
also facilitates observation of KPZ scaling behavior in equal-time
spatial correlations in 2D \cite{Altman_2d_driven_SF_2013}.) The
middle (yellow) curve in Fig.~\ref{Fig. Psi_Psi_temporal_correlation_LogLog_Log_scale_plot}
shows $\bar{C}_{t}(t_{1},t_{2})$ for $\tilde{\gamma}_{l}=1\mathrm{ps}^{-1}$
and a dimensionless linear system size of $L=2^{7}$, corresponding
in physical units to $\sim1.5\times10^{2}\mu\mathrm{m}$. This means
the $Q$ factor is reduced by a factor of $\thicksim30$ compared
to the ones of typical high $Q$ cavities. We note that a decreased
$Q$ factor (also indicating an increased noise strength) can destroy
the condensate if it is smaller than a threshold value. This is because
the reduced cavity lifetime is accompanied by an increased noise level
(cf. Eq. (\ref{eq:SCGLE}) and the subsequent discussion), which acts
to destroy the condensate. However, we remark here that in the simulation
results for this decreased $Q$ factor presented in the following,
the system is still condensed, i.e. we are still working in a low
noise level regime. In addition to the increase of $\tilde{\gamma}_{l}$,
for this simulation we chose a larger value of 6 for the dimensionless
prefactor in $u_{d}$ in Eq. (\ref{eq:params_Exp_rescaled_SCGLE})
instead of $\sim1$ which we obtain for the parameters given in Eq.
(\ref{eq:Exp_parameters_values_decent_cavity}). This choice magnifies
the effective KPZ non-linearity and corresponds to a moderate variation
of the experimental parameters only. In fact, the latter are often
determined only indirectly via fitting simulations to experimental
measurements, and are thus not known with very high precision. In
this setting, the exponent of $\beta=0.317$ obtained from the middle
(yellow) curve in Fig.~\ref{Fig. Psi_Psi_temporal_correlation_LogLog_Log_scale_plot}
indicates that it is promising to search for signatures of KPZ physics
in the first-order temporal coherence of 1D exciton-polariton systems
when the lifetime of polaritons is rather short, so that the intrinsic
non-equilibrium nature is strongly pronounced.

\section{Conclusions and Outlook}

We investigated scaling properties of the long-wavelength dynamics
of 1D driven-dissipative condensate via direct numerical simulations
of the SCGLE, and numerically established the connection to 1D KPZ
universality. We further numerically confirmed the experimental observability
of the non-equilibrium scaling properties of the first order temporal
coherence within the typical current experimental setups of exciton-polariton
condensates if cavities with a reduced $Q$ factor are used. Similar
investigations will be extended to higher dimensions in the future.
Moreover, it is intriguing to investigate the dynamics of the driven-dissipative
condensates at higher noise level, where in particular phase defects,
e.g. phase slips in 1D or vortices in 2D, are expected to play a role
in determining the long-wave length scaling properties of the system's
dynamics. 
\begin{acknowledgments}
We thank I. Boettcher for useful discussions. This work was supported
by the Austrian Science Fund (FWF) through the START grant Y 581-N16,
the SFB FoQuS (FWF Project No. F4006-N16), the Austrian Ministry of
Science BMWF as part of the UniInfrastrukturprogramm of the Focal
Point Scientific Computing at the University of Innsbruck, by German
Research Foundation (DFG) through ZUK 64, and by European Research
Council Synergy Grant UQUAM, the Israel Science Foundation (E.A.). 
\end{acknowledgments}
\appendix

\section{Extraction of $\alpha,\beta,$ and $t_{c}$}

In this appendix we present a more precise determination of the static
roughness exponent $\alpha$ and the dynamical growth exponent $\beta$,
and describe how the crossover time scale $t_{c}$ is extracted in
numerical simulations.

\subsection{Static roughness exponent $\alpha$}

\label{sec:stat-roughn-expon}

\begin{figure}
\includegraphics[width=2.8in]{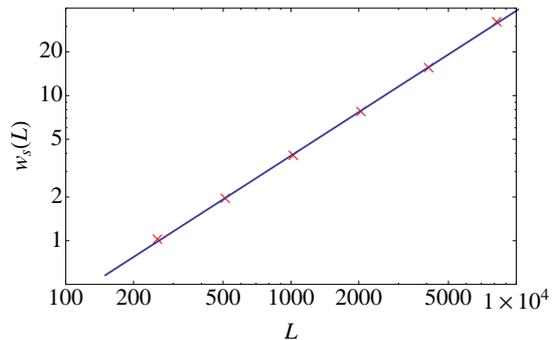}\protect\protect\caption{(Color online) Finite size scaling of $w_{s}(L)$. Points marked by
``$\times$'' denote the numerical value of $w_{s}(L)$ for system
sizes $L=2^{8},2^{9},2^{10},2^{11},2^{12},2^{13}$. The blue line
is a linear fit to the data on the log-log scale, from which we extract
the roughness exponent $\alpha=0.499$. This is in good agreement
with the roughness exponent $\alpha_{\mathrm{KPZ}}$ of the KPZ dynamics
in 1D, $\alpha_{\mathrm{KPZ}}=1/2$. Values of parameters used in
the simulations are $r_{c}=-0.1,u_{c}=0.1,\sigma=0.1,K_{c}=3.0$. }

\label{Fig. roughness_exponent} 
\end{figure}

We extract $\alpha$ from the finite-size scaling of $w_{s}(L)$.
For given system size $L$, we monitor the value of $w(L,t)$ during
a simulation and wait until it reaches a stable value up to statistical
fluctuations at the saturation time $T_{s}$. After $T_{s}$, we continue
simulating the dynamics to the final time point $T_{f}$ with $T_{f}-T_{s}$
at least two times larger than $T_{s}$. Afterwards $w_{s}(L)$ is
extracted according to $w_{s}(L)=(T_{f}-T_{s})^{-1}\int_{T_{s}}^{T_{f}}dt\, w(L,t)$.
In Fig.~\ref{Fig. roughness_exponent} we show the finite size scaling
of $w_{s}(L)$ from the direct simulations of the SCGLE. The extracted
roughness exponent is $\alpha=0.499$, which is in good agreement
with the roughness exponent $\alpha_{\mathrm{KPZ}}$ of the KPZ dynamics
being $\alpha_{\mathrm{KPZ}}=1/2$ in 1D \cite{Halpin-Healy_KPZ_review}.

\subsection{Dynamical growth exponent $\beta$ and crossover time scale $t_{c}$}

\label{sec:dynam-growth-expon}

We extract $\beta$ from the time dependent roughness function $w(L,t)$.
As pointed before, this exponent is related to the dynamical exponent
$z$ and the roughness exponent $\alpha$ via the relation $\beta=\alpha/z$.
Its value is expected to be $1/3$ and $1/4$ for effective KPZ and
EW dynamics, respectively \cite{Halpin-Healy_KPZ_review}.

\begin{figure}
\includegraphics[width=3in]{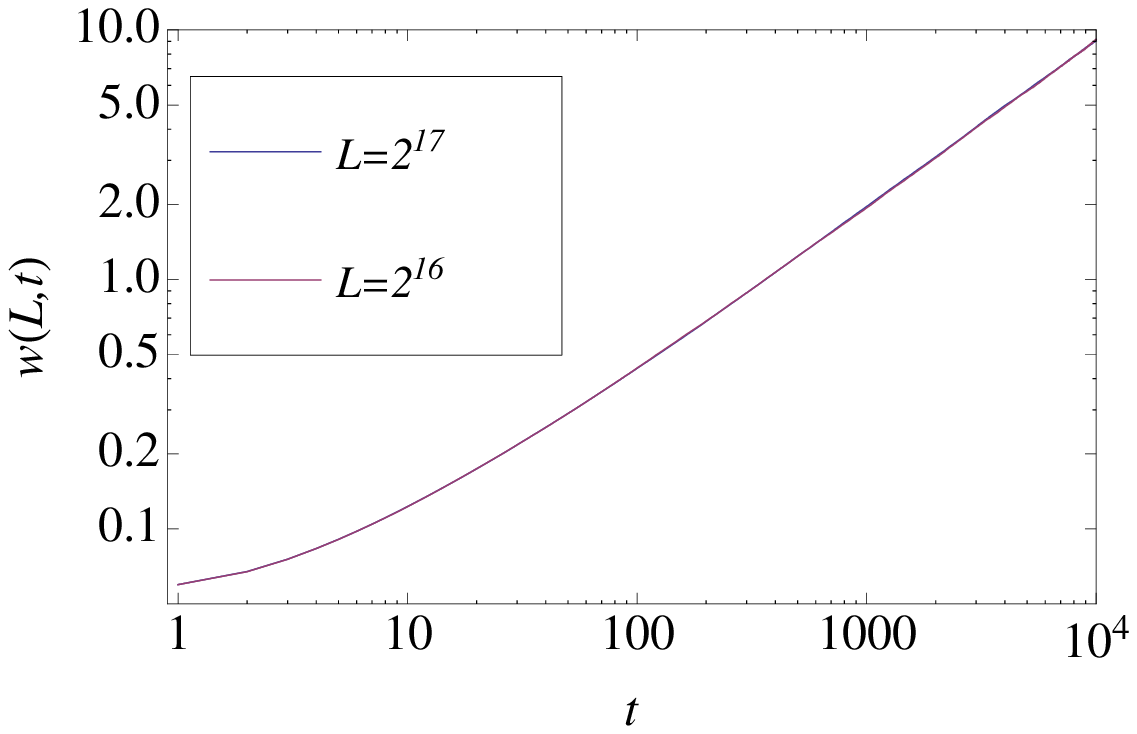}

\includegraphics[width=2.9in]{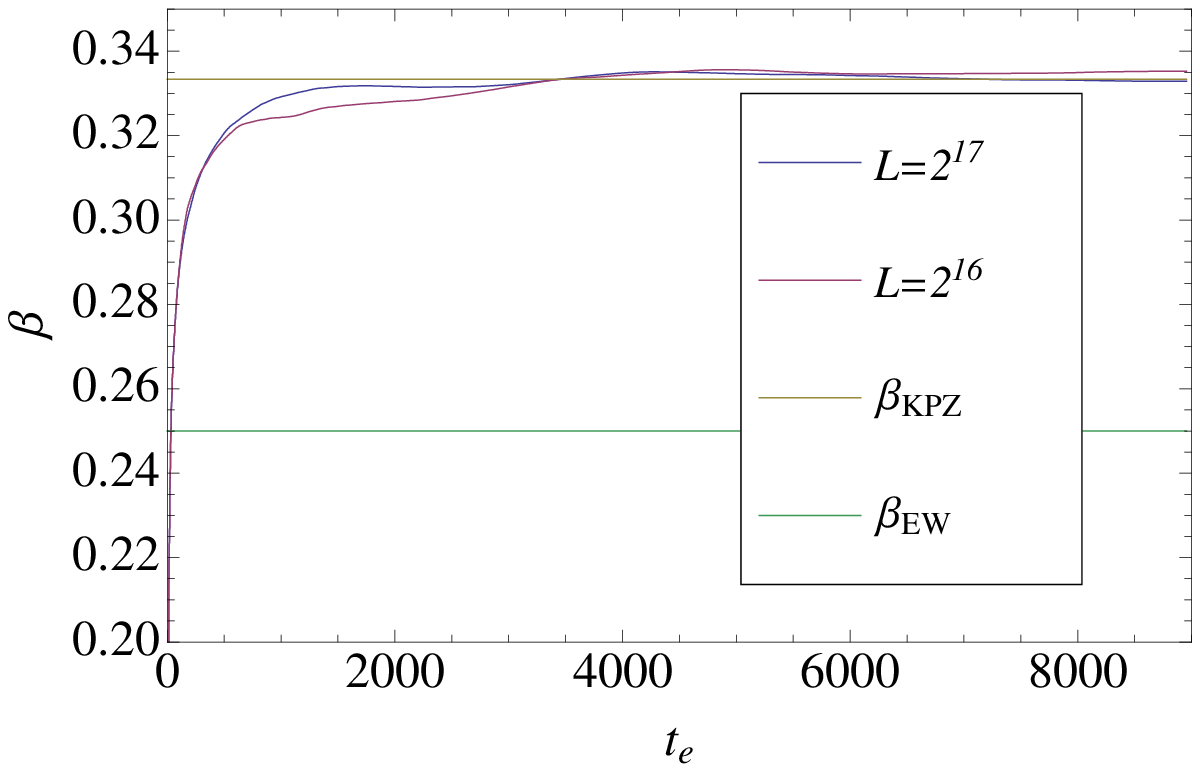}\protect\protect\caption{(Color online) Upper panel: Time dependence of $w(L,t)$ for different
system sizes $L=2^{16},2^{17}$ with $r_{c}=-0.1,u_{c}=0.1,\sigma=0.1,K_{c}=3.0$,
on logarithmic scales. Lower panel: $t_{e}$ dependence of extracted
$\beta$ at different system sizes $L=2^{16},2^{17}$. The growth
exponent of EW dynamics $\beta_{\mathrm{EW}}$ and KPZ dynamics $\beta_{\mathrm{KPZ}}$
are indicated by two lines in the plot to facilitate direct comparison.
From the results at system size $L=2^{17}$, we obtain the dynamical
exponent $\beta=0.335$ from the maximum value of $\beta(t_{e})$,
which is in good agreement with $\beta_{\mathrm{KPZ}}=1/3$.}

\label{Fig. growth_exponnet} 
\end{figure}

In order to reliably extract the exponent $\beta$ it is important
to note that $w(L,t)\propto t^{2\beta}$ is reached only after the
initial crossover time scale $t_{c}$ discussed in Sec.~\ref{sub:Crossover-time-scale}.
In practice we fit $w(L,t)$ to a power law over a long time window
$t\in[t_{e},t_{e}+T]$. We identify the asymptotic scaling by observing
how the exponent depends on the lower cutoff time $t_{e}$. As shown
in Fig.~\ref{Fig. growth_exponnet}, the fitted exponent $\beta$
first grows with $t_{e}$ but then rapidly reaches a plateau. The
value at this plateau represents the asymptotic scaling behavior.
We note however that for this scheme to reflect the KPZ scaling, the
upper cutoff time $t_{e}+T$ should not reach the finite size saturation
time $T_{s}\sim L^{z}$ of the roughness function. If it does, then
we expect the extracted exponent $\beta$ to start decreasing again.
Thus we extract $\beta$ from the maximum value of the fitted exponent
$\beta(t_{e})$. This gives the estimate $\beta=0.335$ consistent
with KPZ dynamics, for which $\beta=1/3$ .

To extract the crossover time $t_{c}$ from the simulations we use
the following scheme. At given system size $L$, we fit the time dependent
roughness function $w(L,t)$ to a double scaling function $c_{\mathrm{EW}}t^{1/2}+c_{\mathrm{KPZ}}t^{2/3}$
in a time interval that extends from zero until a final time $t_{f}$
well before the finite system size effects set in, i.e. $t_{f}\ll T_{s}$.
We then identify $t_{c}$ as the time point where the two scaling
functions have the same contribution to the roughness function, i.e.
$c_{\mathrm{EW}}t_{c}^{1/2}=c_{\mathrm{KPZ}}t_{c}^{2/3}$, giving
rise to $t_{c}=(c_{\mathrm{EW}}/c_{\mathrm{KPZ}})^{6}$.

\end{document}